\documentclass[a4paper]{article}

\usepackage{main}
\usepackage[dvipdfmx]{hyperref}

\title{Singing voice synthesis based on convolutional neural networks}
\name{Kazuhiro Nakamura$^1$, Kei Hashimoto$^{1,2}$, Keiichiro Oura$^{1,2}$, Yoshihiko Nankaku$^{2}$, \\and Keiichi Tokuda$^{1, 2}$}
\address{
  $^1$Department of Research and Development, Techno-Speech, Inc., Nagoya, Japan\\
  $^2$Department of Computer Science, Nagoya Institute of Technology, Nagoya, Japan}
\email{nkazu@techno-speech.com, {bonanza, uratec, nankaku, tokuda}@sp.nitech.ac.jp}

\begin{document}

\maketitle
\begin{abstract}
The present paper describes a singing voice synthesis based on convolutional neural networks ({CNNs}).
Singing voice synthesis systems based on deep neural networks ({DNN}s) are currently being proposed
and are improving the naturalness of synthesized singing voices.
In these systems, the relationship between musical score feature sequences and acoustic feature sequences extracted from singing voices 
is modeled by {DNN}s.
Then, an acoustic feature sequence of an arbitrary musical score is output in units of frames by the trained {DNN}s,
and a natural trajectory of a singing voice is obtained by using a parameter generation algorithm.
As singing voices contain rich expression, a powerful technique to model them accurately is required.
In the proposed technique, long-term dependencies of singing voices are modeled by {CNNs}.
An acoustic feature sequence is generated in units of segments that consist of long-term frames,
and a natural trajectory is obtained without the parameter generation algorithm.
Experimental results in a subjective listening test show that the proposed architecture can synthesize natural sounding singing voices.
\end{abstract}

\noindent\textbf{Index Terms}: Singing voice synthesis, statistical model, acoustic modeling, convolutional neural network

\section{Introduction}

Deep neural networks ({DNN}s), which are artificial neural networks with many hidden layers, 
are attaining significant improvement in various speech processing areas, e.g.,
speech recognition \cite{hinton-IEEE-2012}, speech synthesis \cite{zen-ICASSP-2013, qian-ICASSP-2014} 
and singing voice synthesis \cite{nishimura-Interspeech-2016}.
In {DNN}-based singing voice synthesis, a {DNN} works as an acoustic model that represents a mapping function from musical score feature sequences 
(e.g., phonetic, note key, and note length feature) to acoustic feature sequences (e.g., spectrum, excitation, and vibrato).
{DNN}-based acoustic models can represent complex dependencies 
between musical score feature sequences and acoustic feature sequences more efficiently than hidden Markov model ({HMM})-based acoustic models \cite{watts-ICASSP-2016}.
Neural networks that can model audio waveforms directly, e.g., {W}ave{N}et \cite{oord-arXiv-2016}, 
{S}ample{RNN} \cite{mehri-arXiv-2016}, {W}ave{RNN} \cite{kalchbrenner-arXiv-2018}, 
{FFTN}et \cite{jin-ICASSP-2018}, and {W}ave{G}low \cite{prenger-arXiv-2018}, are currently being proposed.
Such neural networks are used as vocoders in the speech field and improve the quality of synthesized speech compared to conventional vocoders \cite{tamamori-ICASSP-2017}.
The neural vocoders use acoustic features as inputs.
Therefore, accurately predicting acoustic feature sequences from musical score feature sequences by acoustic models 
is still an important issue for generating high quality speech or singing voices.

One limitation of the feed-forward {DNN}-based acoustic modeling \cite{zen-ICASSP-2013} is that the sequential nature of speech is not considered.
Although there are certainly correlations between consecutive frames in speech data,
the feed-forward {DNN}-based approach assumes that each frame is generated independently.
As a solution, recurrent neural networks ({RNN}s) \cite{robinson-NIPS-1988}, 
especially long short-term memory ({LSTM})-{RNN}s \cite{hochreiter-NeuralComput-1997}, 
provide an elegant way to model speech-like sequential data that embodies short and long-term correlations.
Furthermore, this problem can be relaxed by smoothing predicted acoustic features 
using the speech parameter generation algorithm \cite{tokuda-ICASSP-2000},
which utilizes dynamic features as constraints to generate smooth speech parameter trajectories.
On the other hand, some techniques to incorporate the sequential nature of speech data into an acoustic model itself have been proposed \cite{zen-ICASSP-2015, wang-ICASSP-2017}.

This paper proposes an architecture that converts musical score feature sequences to acoustic feature sequences in units of segments 
by using convolutional neural networks ({CNNs}).
The proposed approach  can capture long-term dependencies of singing voices and can generate natural trajectories 
without the speech parameter generation algorithm \cite{tokuda-ICASSP-2000}.
Furthermore, parallel computation can be applied easily, i.e., the training of {CNN}s and the generation of acoustic features are fast,
because there is no recurrent structure in this architecture.

% 論文の構成
The rest of this paper is organized as follows.
Section~2 gives an overview of the {DNN}-based singing voice synthesis system.
Related work is described in Section~3.
Details of the proposed {CNN}-based singing voice synthesis architecture are described in Section~4.
Experimental results in subjective evaluation are given in Section~5.
The key points are summarized, and future work is mentioned in Section~6.

\section{{DNN}-based singing voice synthesis}
\begin{figure}[t]
\begin{center}
\vspace{0.2cm}
\includegraphics[width=1.0\hsize]{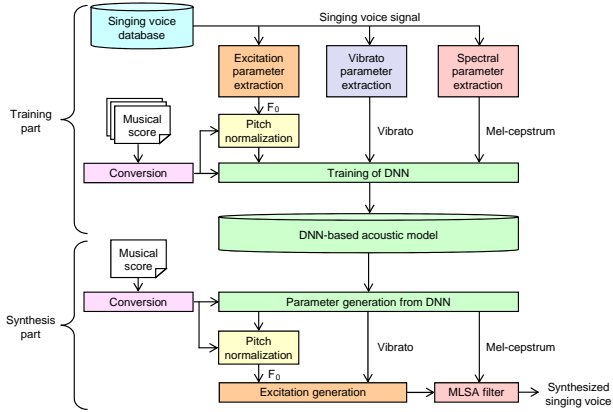}  % TODO: change to DNN system
\end{center}
%\vspace{-0.7cm}
\caption{Overview of {DNN}-based singing voice synthesis system.}
\label{fig:system}
\vspace{-0.4cm}
\end{figure}

A {DNN}-based singing voice synthesis system is quite similar 
to a {DNN}-based text-to-speech synthesis system \cite{zen-ICASSP-2013}.
However, there are distinct differences.
Figure~\ref{fig:system} gives an overview of the {DNN}-based singing voice synthesis system \cite{nishimura-Interspeech-2016, hono-APSIPA-2018}.
It consists of training and synthesis parts.
In the training part, spectrum (e.g., mel-cepstral coefficients), excitation, 
and vibrato parameters are extracted from a singing voice database as acoustic features.
%ビブラートは音高や音量を周期的に揺らす歌唱表現である．
%Vibrato is a singing expression that periodically shakes the pitch.
%Input musical score features and output acoustic features of {DNN}s are time-aligned by well-trained {HMM}s,
Then, musical score feature sequences and acoustic feature sequences are time-aligned by well-trained {HMM}s,
and the mapping between them is modeled by {DNN}s.
In the synthesis part, an arbitrarily given musical score including lyrics to be synthesized is first converted 
into a musical score feature sequence,
and it is mapped to an acoustic feature sequence by the trained {DNN}s.
Next, the speech parameters (spectrum, excitation, and vibrato) are generated 
by a maximum likelihood parameter generation ({MLPG}) algorithm \cite{tokuda-ICASSP-2000}.
It is shown that the quality of the generated speech was improved 
by considering the explicit relationship between static and dynamic features \cite{hashimoto-ICASSP-2015}.
Finally, a singing voice is synthesized from the generated parameters 
by using a vocoder based on a mel log spectrum approximation ({MLSA}) filter \cite{imai-IEICE-1983}.

Rhythm and tempo of music are important factors in singing voice synthesis.
In a human singing voice, there are differences between the start timing of the notes and the singing voices.
The start timing of the singing voice is often earlier than that of a corresponding note.
In order to predict such differences, time-lag models are introduced \cite{hono-APSIPA-2018}.
The naturalness of the synthesized singing voice is improved by accurately predicting the time-lag,
i.e., the start timing and durations of the singing voices.
%Therefore, the start timing of the notes and the durations for each note must be determined 
%from the musical score.
%In a human singing voice, however, there are differences between the start timing of the notes and the singing voices.
%The start timing of the singing voice is often earlier than that of a corresponding note.
%Because this could be an important factor to the naturalness of the synthesized singing voice,
%they are modeled explicitly by time-lag models \cite{hono-APSIPA-2018}.
%However, if the musical score is strictly followed, the synthesized singing voice will be unnatural because of time lags.
%To overcome this problem, the time lags of individual notes are modeled by DNNs \cite{hono-APSIPA-2018}.

Vibrato is one of the important singing techniques that should be modeled,
even though it is not included in the musical score.
Vibrato has been assumed as periodic fluctuations of only $F_0$ for the sake of simplicity,
and it is modeled by sinusoid \cite{yamada-IPS-2009}.
The vibrato $v(\cdot)$ of the $t$ frame in the $i$-th vibrato section $[t_i^{(s)}, t_i^{(e)}]$ can be defined as
\begin{eqnarray}
v\left(m_a (t), m_f (t), i \right) = m_a (t) \sin\left(2 \pi m_f (t) f_s (t - t_i^{(s)}) \right),
\end{eqnarray}
where $m_a (t)$, $m_f (t)$, and $f_s$ correspond to the $F_0$ amplitude of vibrato in cents, the $F_0$ frequency of vibrato in Hz,
and the frame shift, respectively.
Two dimensional parameters, $m_a (t)$ and $m_f (t)$, are added to the acoustic feature vector.

The performance of statistical parametric approaches for singing voice synthesis heavily depends on the training data 
because these approaches are ``corpus-based.''
It is difficult to express contextual factors that barely ever appear in the training data.
Databases including various contextual factors should be used in {DNN}-based singing voice synthesis systems,
whereas it is almost impossible to cover all possible contextual factors because singing voices involve a huge number of them, 
e.g., keys, lyrics, dynamics, note positions, durations, and pitch.
Among them, pitch should be correctly covered because generated $F_0$ trajectories greatly affect the quality of the synthesized singing voices.
To address this problem, a musical-note-level pitch normalization technique has been proposed for {DNN}-based singing voice synthesis systems \cite{nishimura-Interspeech-2016}.
In this technique, the differences between the log $F_0$ sequences extracted from waveforms and the pitch of musical notes are modeled.
This technique makes it possible for {DNN}-based singing voice synthesis systems to generate various singing voices including arbitrary pitch.
Another problem for modeling differences in log $F_0$ is 
how to model log $F_0$ of singing voices including unvoiced frames and musical scores including musical rests.
In \cite{nishimura-Interspeech-2016}, all unvoiced frames and musical rests in musical scores are linearly interpolated and modeled as voiced frames.
%Pitchpadaptive training (PAT) \cite{pitch_adaptive} is used to generate singing voices in any pitch. 

%{DNN}に基づく歌声合成(DNN歌声合成)\cite{hono-APSIPA-2018}の学習部では，楽譜特徴量と音響特徴量の対応関係を学習する．
%音響特徴量としては，テキスト音声合成に用いられるスペクトル，基本周波数，非周期成分に加え，
%音高や音量を周期的に揺らす歌唱表現であるビブラートを表す特徴量が用いられる．
%{DNN}歌声合成は統計的手法であるため，学習データに含まれない音高を合成することは困難である．
%そこで，歌声の対数基本周波数を直接モデル化するのではなく，楽譜情報における音符の音高との差分をモデル化する音高正規化学習が行われる．
%また，楽曲のテンポやリズムに従って歌声を合成する必要があるため，楽譜を元に音符の開始時間や音素継続長を求める必要がある．
%しかし，前ノリ，後ノリ，タメといった発声タイミングに関する歌唱表現や，歌唱者が意識しない発声タイミングの癖により，
%楽譜から計算される音符のタイミングと実際の発声タイミングは必ずしも一致しない．
%そのため，発声タイミングのずれをモデル化することで，自然な発声タイミングの歌声を合成することができる．
%合成部では，合成したい曲の楽譜から抽出した特徴量を{DNN}に与えることで，
%音響特徴量の静的・動的特徴量を推定し，静的・動的特徴量の関係を考慮したパラメータ生成(Maximum likelihood parameter generation; {MLPG})\cite{tokuda-ICASSP-2000}を行うことで，
%滑らかな静的特徴量系列を得る．
%そして，生成された静的特徴量をボコーダへ入力することで歌声を合成することができる．

\section{Related work}
\subsection{Modeling long-term dependencies of speech}

The simplest way to apply neural networks to statistical parametric speech synthesis ({SPSS}) \cite{zen-SCommu-2009} is to use a feed-forward neural network (FFNN) \cite{zen-ICASSP-2013}
as a deep regression model to map linguistic features directly to acoustic features.
One limitation of this architecture is that the mapping between linguistic and acoustic features is one-to-one.
%Though Recurrent Neural Network ({RNN}) internally considers the dependencies on history of future context
%when propagating information, it was limited to a short range.
{RNN}s \cite{robinson-NIPS-1988} provide an elegant way to model speech-like sequential data 
that embody correlations between neighboring frames.
That is, previous input features can be used to predict output features at each frame.
{LSTM-RNN}s \cite{hochreiter-NeuralComput-1997}, which can capture long-term dependencies, 
have been applied to acoustic modeling for {SPSS}.
Fan {\it et al.} and Fernandez {\it et al.} applied deep bidirectional {LSTM-RNN}s, 
which can access input features at both past and future frames, 
to acoustic modeling for SPSS and reported improved naturalness \cite{fan-Interspeech-2014, fernandez-Interspeech-2014}.
%最近では{DNN}音声合成の分野でも，End-to-Endでテキスト等を音声波形に直接変換するようなモデル構造が盛んに研究されているが，
%精度や計算リソースの観点から，実用化は未だ困難な場合が多い．
%一方で，テキスト等から抽出した言語特徴量を音声波形から抽出されたスペクトルや基本周波数などの音響特徴量に変換し，
%ボコーダにより音声波形を生成する手法は，
%実現が比較的容易で高い精度が得られることから，以前より広く研究されている．
%
%音声は可変長の信号系列であり，逐次的に推定を行うことが適切であると考えられてきた．
%これまでに，Feed-Forward NN ({FFNN})をはじめとして，{DNN}からフレーム毎に音響特徴量を出力する手法が提案されている．
%トラジェクトリ学習
%\cite{hashimoto-ICASSP-2016}
%\cite{hono-APSIPA-2018}を行う手法が提案され，
%時間による変動を考慮することで合成音声の品質が改善された．
%{FFNN}はフレーム間が独立したモデル構造となっており，フレーム間の相関を考慮することができないため，これを改善するためにリカレント構造を持つモデルが提案されてきた．
%リカレント構造を持ったモデルとしては，最初にRecurrent Neural Network(RNN)を用いた手法\cite{?}が提案された．
%しかし，{RNN}は記憶できる期間が短いため，長い期間記憶が可能なLong Short-Term Memory Recurrent Neural Network({LSTM-RNN})\cite{fan-Interspeech-2014}が提案され，高い性能を示している．
%また，双方向のリカレント構造を持ったBidirectionarl {LSTM} ({BLSTM})\cite{?}も提案されている．
%また，これまでは静的・動的特徴量をモデルから出力して，それらの尤度をフレーム単位で最大化するように学習を行っていた． % TODO: 数式必要？
%これに対して，系列全体の尤度を最大化するトラジェクトリ学習手法\cite{hashimoto-ICASSP-2016}が提案され，歌声に適用する研究も行われている(\cite{hono-APSIPA-2018})．
Trajectory training is another approach for capturing long-term dependencies of speech.
In {DNN}-based systems, although the frame-level objective function is usually used for {DNN} training,
the sequence-level objective function is used for parameter generation.
To address this inconsistency between training and synthesis, a trajectory training method was introduced into the training process of {DNN}s \cite{hashimoto-ICASSP-2016}.
It was also applied to a singing voice synthesis framework \cite{hono-APSIPA-2018}.

The {RNN}-based systems have the problem of taking time since parallelizing model training and parameter generation is difficult.
And the trajectory training method has the problem that the computational cost increases significantly as the sequence length increases.

\subsection{Acoustic model considering sequential nature of speech}
%{DNN}から出力される音声パラメータ系列は時間方向に不連続になり得るため，
%一般的に，静的特徴量に加えて前後の音響特徴量との関係を表す動的特徴量も同時にモデル化し，
%{MLPG}によるパラメータ生成が行われてきた．
One limitation of the {DNN}-based acoustic modeling is that the sequential nature of speech is not expressed enough.
%Although certainly there are correlations between consecutive frames in speech data,
Although this problem can be relaxed by smoothing predicted acoustic features using the speech parameter generation algorithm \cite{tokuda-ICASSP-2000},
which utilizes dynamic features as constraints to generate smooth trajectories.
However, it is desirable to incorporate the sequential nature of speech data into the acoustic model itself.
Fan {\it et al.} claimed that deep bidirectional {LSTM-RNN}s can generate smooth speech parameter trajectories; thus, no smoothing step was required, 
whereas Zen {\it et al.} reported that having the smoothing step was still helpful with unidirectional {LSTM-RNN}s \cite{zen-ICASSP-2013}.
%また，トラジェクトリ学習を行った場合も，合成時は静的・動的特徴量を出力してMLPGを行うのが一般的である．

As many text-to-speech ({TTS}) applications require fast and low-latency speech synthesis,
an existing problem is the high-latency that the {MLPG} algorithm brings during generation.
An efficient way to remove this problem is not to use the dynamic features during modeling.
Zen {\it et al.} proposed a recurrent output layer \cite{zen-ICASSP-2015},
and Wang {\it et al.} proposed a convolutional output layer \cite{wang-ICASSP-2017}
to achieve smooth transitions between consecutive frames, and accordingly, the {MLPG} is replaced.
%Wang {\it et al.} proposed a conventional output layer based on unidirectional {LSTM} \cite{wang-ICASSP-2017} for high-performance speech synthesis.
%Since it is not straihgtforward to perform low latency 
%パラメータ生成のための層をモデルに追加することで，{MLPG}を行うことなく滑らかなパラメータ系列を生成する手法が提案されている．
%例として，
%The streaming synthesis architecture using undirectional {LSTM-RNNs} with a recurrent output layer which encourages smooth transition between consecutive acoustic frames was proposed. {zen-ICASSP-2015}.
%In these mathods, the parameter generation is integrated into {DNN}s for acoustic modeling and achive low-latency speech synthesis while the quality of is keeped.
They were used with unidirectional {LSTM} to achieve both natural sounding speech and low-latency speech synthesis.
%合成音声の精度を維持しながらレスポンスが改善されたことが報告されている．
%また，数フレームの出力を束ねる畳み込み層を追加する手法\cite{wang-ICASSP-2017}も提案されており，合成音声の精度を改善しつつレスポンスも改善されたことが報告されている．
%このように，{DNN}から直接滑らかな音声パラメータ系列を生成する研究も行われている．

%\vspace{-1mm}
\section{{CNN}-based singing voice synthesis}
%\vspace{-1mm}
\begin{figure*}[t]
  \vspace{-3mm}
  \centering
\includegraphics[width=0.8\hsize]{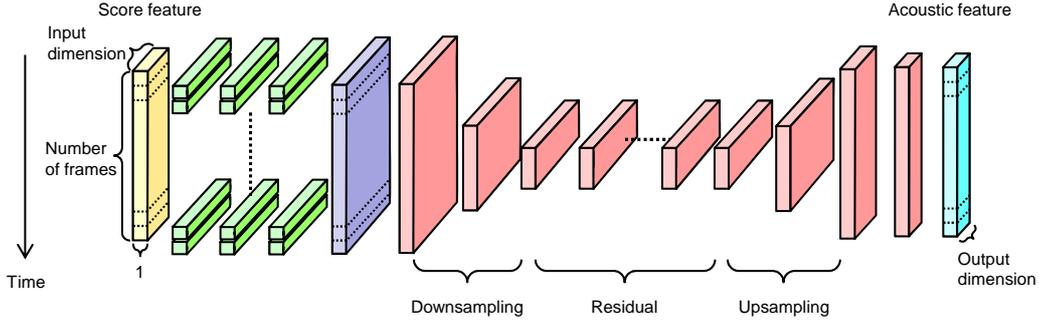}
%  \vspace{-8mm}
  \caption{Overview of proposed {CNN}-based method.}
  \label{fig:CNN-SVSS}
%  \vspace{-3mm}
\end{figure*}

%This section describes the {CNN}-based architecture and loss function for training to generate smooth parameter sequence.

\subsection{{CNN}-based architecture for capturing long-term dependencies of singing voice}
In the proposed method, the relatively long musical score feature sequence, equivalent to several seconds to tens of seconds,
is regarded as a segment and converted to the acoustic feature sequence by {CNNs} at the same time.
The overview of the proposed method is shown in Figure~\ref{fig:CNN-SVSS}.
The first part consists of $1 \times 1$ convolutional layers that convert the musical score feature sequence frame-by-frame.
The dropout technique is used to keep the robustness against the unknown musical scores.
The second part consists of $1 \times n$ convolutional layers,
 and the intermediate output feature sequence of the first part is converted 
to the acoustic feature sequence in units of segments.
The dimension of the acoustic features is represented as the number of channels of the output features.
The size of the segment is $1 \times T$, and $T$ means the number of frames of each segment.
Since a fully convolutional network ({FCN}) \cite{long-CVPR-2015} is used as the {CNN} structure, the segment size $T$ is adjustable.
These parts are integrated and trained simultaneously.

The relationship between a musical score feature vector sequence $\bm{s}=[\bm{s}_1^{\top}, \bm{s}_2^{\top}, \dots, \bm{s}_T^{\top}]^{\top}$
and an acoustic feature vector sequence $\bm{c}=[\bm{c}_1^{\top}, \bm{c}_2^{\top}, \dots, \bm{c}_T^{\top}]^{\top}$
is represented as follows
%\vspace{-2mm}
\begin{eqnarray}
%\bm{c} = G\left([F\left(\bm{s}_1\right)^{\top}, F\left(\bm{s}_2\right)^{\top}, \dots, F\left(\bm{s}_T\right)^{\top}]^{\top}\right)
\bm{c} = G([F(\bm{s}_1)^{\top}, F(\bm{s}_2)^{\top}, \dots, F(\bm{s}_T)^{\top}]^{\top}),
\end{eqnarray}
where $F(\cdot)$ is a frame-by-frame mapping function in the first part,
and $G(\cdot)$ is a segment-by-segment mapping function in the second part.

As the pitch of musical notes greatly affects the synthesized singing voices,
we concatenate them with the output features from the first part of the proposed {CNNs} and use them as the input of the second part.
In particular, the alignment of notes is adjusted to the recorded singing voices,
and log $F_0$ parameters from musical notes are concatenated.
The musical rests in musical scores are interpolated linearly.
The effectiveness of the use of log $F_0$ parameters extracted from the interpolated musical scores was confirmed in the preliminary subjective experiment.

\subsection{Loss function for obtaining smooth parameter sequence without parameter generation algorithm}
\begin{figure}[t]
%  \vspace{-3mm}
  \centering
  \includegraphics[width=\linewidth]{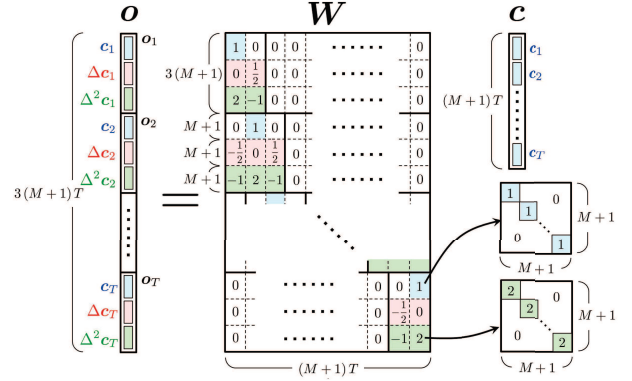}
%  \vspace{-8mm}
  \caption{Example of relationship between static feature vector sequence $\bm{c}$ and singing voice parameter vector sequence $\bm{o}$ in a matrix form.}
  \label{fig:window}
%  \vspace{-3mm}
\end{figure}

In the proposed method, a loss function based on the likelihood of $\bm{o}_t$ is used to obtain smooth acoustic feature sequences.
A parameter vector $\bm{o}_t$ of a singing voice consists of a $D$-dimensional static feature vector
$\bm{c}_t = \left[c_t (1), c_t (2), \dots, c_t (D) \right]^\top$
and their dynamic feature vectors $\Delta^{(\cdot)} \bm{c}_t$.
%\vspace{-1mm}
\begin{eqnarray}
\bm{o}_t = [\bm{c}_t^\top, \Delta^{(1)} \bm{c}_t^\top, \Delta^{(2)} \bm{c}_t^\top ]^\top
\end{eqnarray}
The sequences of the singing voice parameter vectors and the static feature vectors can be written in vector forms as follows
\begin{eqnarray}
\bm{o} &=& [\bm{o}_t^\top, \dots, \bm{o}_t, \dots, \bm{o}_T^\top]^\top \\
\bm{c} &=& [\bm{c}_t^\top, \dots, \bm{c}_t, \dots, \bm{c}_T^\top]^\top,
\end{eqnarray}
where $T$ is the number of frames.
The relation between $\bm{o}$ and $\bm{c}$ can be represented by $\bm{o} = \bm{W}\bm{c}$,
where $\bm{W}$ is a window matrix extending the static feature vector sequence $\bm{c}$ 
to the singing voice parameter vector sequence $\bm{o}$ (Fig.~\ref{fig:window}).

In the training part, an objective function is defined as
\begin{eqnarray}
\mathcal{L} = \mathcal{N}\left(\bar{\bm{o}} | \bm{o} , \bm{\Sigma}\right),
\end{eqnarray}
where $\bar{\bm{o}}$ is represented by $\bar{\bm{o}} = \bm{W}\bar{\bm{c}}$,
where $\bar{\bm{c}}$ is the static feature vector sequence of the recorded singing voice.
$\bm{\Sigma}$ is a globally tied covariance matrix given by 
\begin{eqnarray}
\bm{\Sigma} = diag [\bm{\Sigma}_1, \dots, \bm{\Sigma}_t, \dots, \bm{\Sigma}_T]
\end{eqnarray}
and is updated during the training.

The proposed method can generate a natural trajectory without the parameter generation algorithm
by considering not only static features but also dynamic features in the training part of the {CNNs}.

\section{Experiments}
\subsection{Experimental conditions}
% 提案手法の有効性を検証するために，主観評価実験を行った．
To evaluate the effectiveness of the proposed method, a subjective comparison test of mean opinion scores ({MOS}) was conducted.
% 学習用の歌声データベースとして女性1名による童謡55曲とJ-POP 55曲を使用し，
For training, 55 Japanese children's songs and 55 {J-POP} songs by a female singer were used,
%テストデータは学習データに含まれていないJ-POP 5曲とした．
and 5 other {J-POP} songs were used for the test.
% サンプリング周波数は48kHz，分析周期は5msである．
Singing voice signals were sampled at 48 {kHz} and windowed with a 5-ms shift.
The number of quantization bits was 16.
% 音響特徴量としては，STRAIGHT分析によって得られた50次元のメルケプストラム，対数基本周波数，22次元の非周期成分，振幅および周波数から成る2次元のビブラートパラメータを用いた．
The feature vectors consisted of 0-th through 49-th STRAIGHT \cite{kawahara-SpeechCommunication-1999} mel-cepstral coefficients,
log ${F_0}$ values, 22-dimensions aperiodicity measures, and 2-dimensions vibrato parameters.
The vibrato parameter vectors consisted of amplitude (cent) and frequency ({Hz}).
% 対数基本周波数とビブラートパラメータに関しては，値を持たない区間を線形補間し，
% 値の有無を表すパラメータをそれぞれに加え，77次元の音響特徴量ベクトルとした．
The areas that do not have a value were interpolated linearly for log ${F_0}$ and vibrato parameters,
and with/without value flags were added to feature vectors.
%入力特徴量には，724次元の楽譜情報および状態位置情報，状態継続長情報，状態内フレーム位置情報を用いた．(連続値：108+14)
The input features including 724 binary features for categorical contexts (e.g., the current phoneme identity and the key of the current measure) 
and 122 numerical features for numerical contexts (e.g., the number of phonemes in the current syllable and the absolute pitch of the current musical note) were used.
% 最小値が0，最大値が1となるように正規化して用いた．
Both input and output features in the training data for the {DNN}s were normalized;
the input features were normalized to be within 0.00--1.00, %on the basis of their minimmum and maximum values in the training data, 
and the output features were normalized to be within 0.01--0.99 on the basis of their minimum and maximum values in the training data.

% 学習データのアライメント
Five-state, left-to-right, no-skip hidden semi-Markov models ({HSMMs}) were used to obtain the time alignment of acoustic features 
in state for training the {DNN}-based acoustic models.
% 状態継続長情報については，学習時にはHMMを用いた状態アライメントから，合成時には別途学習した{FFNN}の出力から求めた．
And the state duration was predicted by {FFNNs} trained from the time alignment of training data.

The FFNN-based singing voice synthesis was used as the conventional method.
The conventional system had 3 layers with 2048 units, and dropout with probability 0.2 was used.
The acoustic features and their dynamic features (delta and delta-delta) were output,
and the {MLPG} algorithm was used to obtain the smooth feature sequences.
In the proposed system, the first part had the same structure as the conventional system.
The second part had 2 layers for down-sampling, 9 layers that have residual structure, and 2 layers for up-sampling.
The data were separated into segments of 2000 frames and used for training and generation,
and 100 adjacent frames were cross-faded in the generation step.
In both systems, the ReLU activation function was used for hidden layers, and the sigmoid one was used for the output layer.

A {MLSA}-based vocoder \cite{imai-IEICE-1983} and a {W}ave{N}et vocoder were used
for conversion from acoustic feature sequences to singing voice waveforms.
% {WaveNet}ボコーダは，収録音声を$\mu$-lawアルゴリズムを用いて8bitに量子化した波形から学習した．
The singing voice signals to train {W}ave{N}et were sampled at 48 kHz and quantized from 16 bits to 8 bits by using the $\mu$-law quantizer \cite{PCM}.
Mel-cepstrum-based noise shaping and prefiltering were applied to the quantization step \cite{yoshimura-IEEE-2018}.
% 強度を調整するパラメータの設定
The parameters for adjusting the intensity in the noise shaping and prefiltering were set as $\gamma=0.4$，$\beta=0.2$.
% {WaveNet}の構造には，dilationを1,2,3,...,512とした10層からなるブロックを3段重ねたものを用いた．
The dilations of the WaveNet model were set to $1, 2, 4, \dots, 512$.
Ten dilation layers were stacked three times.
% WaveNetのチャネル数
The sizes of the channels for dilations, residual blocks, and skip-connections were 256, 512, and 256, respectively.

\subsection{Experimental results}
\label{sec:subjective}
%提案手法の有効性を評価するため，主観評価実験を行った．
%上記4手法の合成音声を，自然性に関する5段階のMOS試験によって評価した．
The 5-points {MOS} evaluation (5: natural -- 1: poor) for the naturalness was conducted.
%被験者は15人であり，各被験者はテストデータ5曲からランダムに選択された各手法10フレーズずつを評価した．
Fifteen subjects evaluated ten phrases randomly selected from the test data.

\begin{figure}[t]
  \centering
  \vspace{3mm}
  \includegraphics[width=\linewidth]{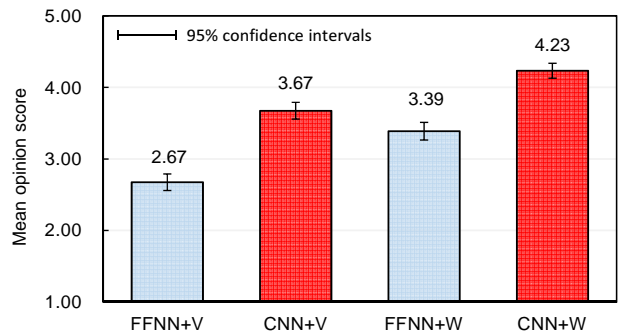}
%  \vspace{-4mm}
  \caption{Subjective experimental results.}
  \label{fig:MOS}
\end{figure}

%図\ref{fig:MOS}に主観評価実験の結果を示す.
Figure~\ref{fig:MOS} shows the results of the MOS evaluation.
\textbf{FFNN+V} and \textbf{FFNN+W} represent conventional systems,
and \textbf{CNN+V} and \textbf{CNN+W} represent proposed systems.
\textbf{V} and \textbf{W} mean {MLSA}-based vocoder and {W}ave{N}et vocoder, respectively.
%\textbf{FFNN}は従来の{FFNN}に基づくフレーム単位の音響特徴量生成，
%\textbf{CNN}は{CNN}に基づくセグメント単位の音響特徴量生成である．
%また，\textbf{V}と\textbf{W}はそれぞれ{MLSA}フィルタによるボコーダと{WaveNet}ボコーダによる音声波形生成を表している．

%図より，\textbf{CNN+V}，\textbf{CNN+W}は，従来手法からスコアを大きく改善した．
The {CNN}-based proposed systems (\textbf{CNN+V} and \textbf{CNN+W}) 
outperformed the {FFNN}-based conventional systems (\textbf{FFNN+V} and \textbf{FFNN+W})
as shown in Figure~\ref{fig:MOS}.
%{CNN}を用いて音声の時間による変動をモデル化することで，合成音声の自然性が向上したことがわかる．
These results indicate that the naturalness of the synthesized singing voice is drastically improved by modeling the time-dependent variation by {CNN}s. 
%また，\textbf{FFNN}，\textbf{CNN}の両手法において，{WaveNet}ボコーダは，{MLSA}フィルタによるボコーダと比較し高いスコアを示した．
And the {W}ave{N}et vocoder (\textbf{FFNN+W} and \textbf{CNN+W}) showed a better score than the {MLSA}-based vocoder
(\textbf{FFNN+V} and \textbf{CNN+V}), respectively.

\section{Conclusions}

% 本稿ではCNNに基づく歌声合成手法を提案した．
In this paper, a {CNN}-based acoustic modeling technique has been proposed for singing voice synthesis.
% CNNにより音声の時間による変動をモデル化し，楽譜から抽出した特徴量の時系列から音響特徴量の時系列へセグメント単位で変換する手法を提案した．
Long-term dependencies of singing voices that contain rich expressions were modeled by {CNNs}.
Musical score feature sequences from musical scores were converted to acoustic feature sequences in units of segments,
and natural speech parameter trajectories were obtained without the conventional speech parameter generation algorithm.
Furthermore, parallel computation can be applied easily %, i.e., the training of {CNN}s and the generation of acoustic features are fast,
because there is no recurrent structure in the proposed architecture.
%主観評価実験により，提案手法が合成音声の自然性を大きく改善することを示した．
Experimental results show that the proposed system gives more natural synthesized singing voices.
% 今後の課題として，正しい音高等の事前情報を付加することで調子外れを補正する手法の検討，テキスト音声合成に適用した場合の評価実験などが挙げられる．
Future work includes correcting out-of-tone utterances in singing voices by using prior distribution of tone,
evaluations of the proposed architecture on TTS,
and parameter tuning for practical use.

%\section{Acknowledgements}
%
%The ISCA Board would like to thank the organizing committees of the past INTERSPEECH conferences for their help and for kindly providing the template files. \\
%Note to authors: Authors should not use logos in acknowledgement section; rather authors should acknowledge corporations by naming them only.
%

\bibliographystyle{IEEEtran}

\bibliography{main}

% Generated by IEEEtran.bst, version: 1.13 (2008/09/30)
\begin{thebibliography}{10}
\providecommand{\url}[1]{#1}
\csname url@samestyle\endcsname
\providecommand{\newblock}{\relax}
\providecommand{\bibinfo}[2]{#2}
\providecommand{\BIBentrySTDinterwordspacing}{\spaceskip=0pt\relax}
\providecommand{\BIBentryALTinterwordstretchfactor}{4}
\providecommand{\BIBentryALTinterwordspacing}{\spaceskip=\fontdimen2\font plus
\BIBentryALTinterwordstretchfactor\fontdimen3\font minus
  \fontdimen4\font\relax}
\providecommand{\BIBforeignlanguage}[2]{{%
\expandafter\ifx\csname l@#1\endcsname\relax
\typeout{** WARNING: IEEEtran.bst: No hyphenation pattern has been}%
\typeout{** loaded for the language `#1'. Using the pattern for}%
\typeout{** the default language instead.}%
\else
\language=\csname l@#1\endcsname
\fi
#2}}
\providecommand{\BIBdecl}{\relax}
\BIBdecl

\bibitem{hinton-IEEE-2012}
G.~Hinton, L.~Deng, D.~Yu, G.~E. Dahl, A.~Mohamed, N.~Jaitly, A.~Senior,
  V.~Vanhoucke, P.~Nguyen, T.~N. Sainath \emph{et~al.}, ``Deep neural networks
  for acoustic modeling in speech recognition: The shared views of four
  research groups,'' \emph{IEEE Signal Processing Magazine}, vol.~29, no.~6,
  pp. 82--97, 2012.

\bibitem{zen-ICASSP-2013}
H.~Zen, A.~Senior, and M.~Schuster, ``Statistical parametric speech synthesis
  using deep neural networks,'' \emph{Proceedings of ICASSP}, pp. 7962--7966,
  2013.

\bibitem{qian-ICASSP-2014}
Y.~Qian, Y.~Fan, W.~Hu, and F.~K. Soong, ``On the training aspects of deep
  neural network ({DNN}) for parametric {TTS} synthesis,'' \emph{Proceedings of
  ICASSP}, pp. 3829--3833, 2014.

\bibitem{nishimura-Interspeech-2016}
M.~Nishimura, K.~Hashimoto, K.~Oura, Y.~Nankaku, and K.~Tokuda, ``Singing voice
  synthesis based on deep neural networks,'' \emph{Proceedings of Interspeech},
  pp. 2478--2482, 2016.

\bibitem{watts-ICASSP-2016}
O.~Watts, G.~E. Henter, T.~Merritt, Z.~Wu, and S.~King, ``From {HMM}s to
  {DNN}s: where do the improvements come from?'' \emph{Proceedings of ICASSP},
  pp. 5505--5509, 2016.

\bibitem{oord-arXiv-2016}
A.~van~den Oord, S.~Dieleman, H.~Zen, K.~Simonyan, O.~Vinyals, A.~Graves,
  N.~Kalchbrenner, A.~W. Senior, and K.~Kavukcuoglu, ``{W}ave{N}et: {A}
  generative model for raw audio,'' \emph{CoRR}, vol. abs/1609.03499, 2016.

\bibitem{mehri-arXiv-2016}
S.~Mehri, K.~Kumar, I.~Gulrajani, R.~Kumar, S.~Jain, J.~Sotelo, A.~Courville,
  and Y.~Bengio, ``{S}ample{RNN}: An unconditional end-to-end neural audio
  generation model,'' \emph{arXiv preprint arXiv:1612.07837}, 2016.

\bibitem{kalchbrenner-arXiv-2018}
N.~Kalchbrenner, E.~Elsen, K.~Simonyan, S.~Noury, N.~Casagrande, E.~Lockhart,
  F.~Stimberg, A.~van~den Oord, S.~Dieleman, and K.~Kavukcuoglu, ``Efficient
  neural audio synthesis,'' \emph{arXiv preprint arXiv:1802.08435}, 2018.

\bibitem{jin-ICASSP-2018}
Z.~Jin, A.~Finkelstein, G.~J. Mysore, and J.~Lu, ``{FFT}net: A real-time
  speaker-dependent neural vocoder,'' \emph{Proceedings of ICASSP}, pp.
  2251--2255, 2018.

\bibitem{prenger-arXiv-2018}
R.~Prenger, R.~Valle, and B.~Catanzaro, ``{W}ave{G}low: A flow-based generative
  network for speech synthesis,'' \emph{arXiv preprint arXiv:1811.00002v1},
  2018.

\bibitem{tamamori-ICASSP-2017}
A.~Tamamori, T.~Hayashi, K.~Kovayashi, K.~Takeda, and T.~Toda,
  ``Speaker-dependent {W}ave{N}et vocoder,'' \emph{Proceedings of ICASSP}, pp.
  1118--1122, 2017.

\bibitem{robinson-NIPS-1988}
A.~Robinson and F.~Fallside, ``Static and dynamic error propagation networks
  with application to speech coding,'' \emph{Proceedings of NIPS}, pp.
  632--641, 1988.

\bibitem{hochreiter-NeuralComput-1997}
S.~Hochreiter and J.~Schmidhuber, ``Long short-term memory,'' \emph{Neural
  Comput.}, pp. 1735--1780, 1997.

\bibitem{tokuda-ICASSP-2000}
K.~Tokuda, T.~Yoshimura, T.~Masuko, T.~Kobayashi, and T.~Kitamura, ``Speech
  parameter generation algorithms for {HMM}-based speech synthesis,''
  \emph{Proceedings of ICASSP}, vol.~3, pp. 1315--1318, 2000.

\bibitem{zen-ICASSP-2015}
H.~Zen and H.~Sak, ``Unidirectional long short-term memory recurrent neural
  network with recurrent output layer for low-latency speech synthesis,''
  \emph{Proceedings of ICASSP}, pp. 4470--4474, 2015.

\bibitem{wang-ICASSP-2017}
W.~Wang and B.~Xu, ``Combining unidirectional long short-term memory with
  convolutional output layer for high-performance speech synthesis,''
  \emph{Proceedings of ICASSP}, pp. 5500--5504, 2017.

\bibitem{hono-APSIPA-2018}
Y.~Hono, S.~Murata, K.~Nakamura, K.~Hashimoto, K.~Oura, Y.~Nankaku, and
  K.~Tokuda, ``Recent development of the {DNN}-based singing voice synthesis
  system - {Sinsy},'' \emph{Proceedings of APSIPA ASC}, pp. 1003--1009, 2018.

\bibitem{hashimoto-ICASSP-2015}
K.~Hashimoto, K.~Oura, Y.~Nankaku, and K.~Tokuda, ``The effect of neural
  networks in statistical parametric speech synthesis,'' \emph{Proceedings of
  ICASSP}, pp. 4455--4459, 2015.

\bibitem{imai-IEICE-1983}
S.~Imai, K.~Sumita, and C.~Furuichi, ``Mel log spectral approximation filter
  for speech synthesis,'' \emph{{IECE} Translations on Fundamentals (Japanese
  Edition)}, vol. J66-A, pp. 122--129, 1983.

\bibitem{yamada-IPS-2009}
T.~Yamada, S.~Muto, Y.~Nankaku, S.~Sako, and K.~Tokuda, ``Vibrato modeling for
  {HMM}-based singing voice synthesis,'' \emph{Proceedings of Information
  Processing Society of Japan}, vol. 2009-NUS-80, no.~5, pp. 1--6, 2009.

\bibitem{zen-SCommu-2009}
H.~Zen, K.~Tokuda, and A.~W. Black, ``Statistical parametric speech
  synthesis,'' \emph{Speech Communication}, vol.~51, no.~11, pp. 1039--1064,
  2009.

\bibitem{fan-Interspeech-2014}
Y.~Fan, Y.~Qian, F.~Xie, and F.~K. Soong, ``{TTS} synthesis with bidirectional
  {LSTM} based recurrent neural networks,'' \emph{Proceeding of Interspeech},
  pp. 964--1968, 2014.

\bibitem{fernandez-Interspeech-2014}
R.~Fernandez, A.~Rendel, B.~Ramabhadren, and R.~Hoory, ``Prosody contour
  prediction with long short-term memory, bidirectional, deep recurrent neural
  networks,'' \emph{Proceedings of Interspeech}, pp. 2268--2272, 2014.

\bibitem{hashimoto-ICASSP-2016}
K.~Hashimoto, K.~Oura, Y.~Nankaku, and K.~Tokuda, ``Trajectory training
  considering global variance for speech synthesis based on neural networks,''
  \emph{Proceedings of ICASSP}, pp. 5600--5604, 2016.

\bibitem{long-CVPR-2015}
J.~Long, E.~Shelhamer, and T.~Darrell, ``Fully convolutional networks for
  semantic segmentation,'' \emph{Proceedings of CVPR}, pp. 3431--3440, 2015.

\bibitem{kawahara-SpeechCommunication-1999}
H.~Kawahara, M.~K. Ikuyo, and A.~Cheveigne, ``Restructuring speech
  representations using the pitch-adaptive time-frequency smoothing and an
  instantaneous-frequency-based f0 extraction: Possible role of a repetitive
  structure in sounds,'' \emph{Speech Communication}, vol.~27, no.~3, pp.
  187--207, 1999.

\bibitem{PCM}
``Pulse code modulation ({PCM}) of voice frequencies,'' \emph{ITU-T
  Recommendation G.711}, 1988.

\bibitem{yoshimura-IEEE-2018}
T.~Yoshimura, K.~Hashimoto, K.~Oura, Y.~Nankaku, and K.~Tokuda,
  ``Mel-cepstrum-based quantization noise shaping applied to
  neural-network-based speech waveform synthesis,'' \emph{IEEE/ACM Transactions
  on Audio, Speech, and Language Processing}, vol.~26, no.~7, pp. 1173--1180,
  2018.

\end{thebibliography}

\end{document}